## High magnetic field response of superconductivity dome in quantum artificial High-$Tc$ superlattices with variable geometry

Gaetano Campi[1,2,*], Andrea Alimenti[3,4,†], Sang-Eon Lee[5,‖], Luis Balicas[5,#], Fedor F. Balakirev[6,¶], G. Alexander Smith[6,§], Gennady Logvenov[7,‡] and Antonio Bianconi[1,2,‖‖]

[1]Institute of Crystallography, National Research Council, CNR, Via Salaria Km 29.3, 00015 Monterotondo Rome, Italy

[2]Rome International Center for Materials Science Superstripes RICMASS, Via dei Sabelli 119A, 00185 Rome, Italy

[3]Department of Industrial, Electronic and Mechanical Engineering, Roma Tre University, Via Vito Volterra 62, 00146 Rome, Italy

[4]Istituto Nazionale di Fisica Nucleare INFN, Sezione Roma Tre, Via della Vasca Navale 84, 00146 Rome, Italy

[5]National High Magnetic Field Laboratory (NHMFL), Florida State University (FSU), Tallahassee, Florida 32310, USA

[6]National High Magnetic Field Laboratory (NHMFL), Los Alamos National Laboratory (LANL), Los Alamos, New Mexico 87545, USA

[7]Max Planck Institute for Solid State Research, Heisenbergstraße 1, 70569 Stuttgart, Germany

Corresponding authors: Gaetano Campi*, Andrea Alimenti†, Antonio Bianconi ‖‖

Email: * gaetano.campi@cnr.it, † andrea.alimenti@uniroma3.it, # balicas@magnet.fsu.edu, ‖ sangeon.lee@fsu.edu,

¶ fedor@lanl.gov, § gasmith@lanl.go, ‡ g.logvenov@fkf.mpg.de, ‖‖ antonio.bianconi@ricmass.eu

### Abstract

It is known that cuprate artificial high Tc superlattices (AHTS) with period d, composed of quantum wells confining interface space charge in stoichiometric Mott insulator layers (S), with thickness L, at the interface with overdoped normal metallic cuprate layers (N) show a superconducting dome by tuning the geometric L over d ratio of the SNSN superlattice with the top predicted by quantum material design engineering quantum size effects. Here we report high-field magneto transport measurements up to 41 Tesla of AHTS across the entire superconducting dome. The results show the universal upward-concave behavior of the temperature dependent upper critical magnetic field in low Tc samples at rising edge and drop edge of the dome providing strong evidence consistent with two-band superconductivity for two-band superconductivity in agreement with multigap theory used for quantum design of the SNSN superlattices. The measured superconducting coherence length demonstrates that atomic-scale engineering controls not only the critical temperature but also the intrinsic pair size at Fano-Feshbach resonances physics paving the way toward next generation quantum devices and shedding light on unconventional superconductivity.

### I. Introduction

Artificial high-$T_C$ superlattices (AHTS) composed of quantum wells of Mott Insulators intercalated by metallic overdoped perovskite units have recently emerged as a paradigmatic platform for exploring unconventional superconductivity with the key role of interface internal local electric field spin orbit coupling [1-5]. AHTS are atomically engineered heterostructures composed of alternating atomic layers of Mott insulator $La_2CuO_4$, (LCO) and overdoped metal $La_{1.55}Sr_{0.45}CuO_4$, (LSCO). Superconductivity emerges specifically within quantum wells of LCO at the LCO/LSCO interfaces due to space-charge reconstruction [1-4]. These systems are realized by precise atomic layer molecular beam epitaxy (MBE) using the Bianconi-Perali-Valletta (BPV) theory [1,3] for engineering unconventional superconductivity which enables the control of fundamental parameters, such as the ratio L/d between the superconducting layer thickness L and the superlattice period d, that directly tune the electronic structure and pairing mechanisms across the superconducting dome [3].

The BPV theory explains high-temperature superconductivity in quantum complex matter made of nanoscale building blocks (such as natural cuprates and hydrides) and in AHTS (Artificial High-Tc Superlattices), which display a multiband/multigap electronic scenario composed of a mixture of superconducting gaps. In these systems, the BPV theory includes the quantum exchange interaction between gaps in the Bardeen-Cooper-Schrieffer (BCS) regime and other gaps in the intermediate BEC-BCS crossover regime between Bose–Einstein condensation (BEC) and BCS pairing. This interplay gives rise to a superconducting dome controlled by a Fano-Feshbach shape resonance associated with pair-exchange transfer between different gaps. It posits that the Fano-Feshbach shape resonance producing the superconducting dome arises from interference between multiple superconducting gaps when the chemical potential is tuned in proximity to a topological Lifshitz transition corresponding to a neck-opening process, which significantly amplifies the critical temperature at the maximum of the dome. This framework has been applied to AHTS [1-5] where the chemical potential is tuned by the variable geometric ratio L/d reaching the Fano-Feshbach shape resonance. This resonance is characterized by destructive interference of the superconducting gaps at the topological transition, with the appearing of a new Fermi surface with very low $T_C$, and predicted a constructive interference of the superconducting gaps at the Lifshitz topological transition from a donut to corrugated cylinder in the new Fermi surface in presence of Rashba spin-orbit coupling, reaching the maximum $T_C$ [1,3,5].

In our previous work [5], focusing on superlattices with geometric ratio L/d near the critical value of 2/3 corresponding to optimal doping, $<\delta> \approx 0.15$, we documented both a remarkable enhancement of the superconducting critical temperature, $T_C$, controlled by Fano-Feshbach shape resonance mechanism [1,3,6,7], coexisting with a linear resistivity regime at higher temperatures, in the normal phase [2,8]. This behavior is expected for samples where the chemical potential is tuned at the Van Hove singularity of the second quantum subband [1,3]. A hallmark of this multi-gap superconductivity is the distinctive behavior of the temperature dependent upper critical field $\mu_0H_{c2}(T)$. Contrary to the canonical single-band model prediction, which prescribes a convex, downward-bending for the $\mu_0H_{c2}(T)$ curve as a function of temperature, our measurements on optimally tuned superlattices revealed an anomalous, concave-up profile [5]. Such a deviation from single-band behavior [9] is a recognized fingerprint of two-band (and in general multigap) superconductivity, previously observed in $MgB_2$ [10,11] and interfacial $SrTiO_3$ [12-14]. As corroborated by multiple contemporaneous studies [15-29], the upward curvature and scaling of $\mu_0H_{c2}(T)$ provides a quantitative probe for the presence of multiple condensates with distinct Fermi velocities and pairing characteristics [30-35].

The present work extends this systematic investigation to a broader set of superlattices, encompassing both the samples on the rising edge of the dome L/d=0.44, 0.50 and on the drop edge of the dome L/d=0.875, 0.89 regions of the phase diagram, thus enabling a comprehensive mapping of the interplay between nanostructure, quantum resonance effects, and the evolution of multiband superconductivity across the superconducting dome. High magnetic field transport experiments up to 41 T were performed at the National High Magnetic Field Laboratory (NHMFL, Tallahassee, FL), with a focus on extracting upper critical magnetic fields $\mu_0H_{c2}(T)$ and the underlying coherence lengths $\xi$. Our primary aim is the direct experimental validation of the theoretical prediction that two-band superconductivity with distinct Fermi velocities is not limited to optimal doping, but persists throughout the dome, shaped by the interplay between quantum geometry and band structure quantum engineering [1-7].

The anomalous temperature dependence of the upper critical field in these artificial heterostructures at the rising edge and drop edge of the dome as at the top of the dome [5] unveils the physics of shape resonances in the BCS-BEC crossover [3,6] in engineered artificial High $T_C$ superlattices. Our results not only confirm the universality of multigap phenomenology in oxide superlattices, but also demonstrate the tunability of superconducting properties via atomic-scale design, holding promise for the development of the next-generation high-$T_C$ materials.

## II. Results

We have systematically investigated the resistivity as a function of temperature and applied magnetic fields for a new series of artificial high- $T_C$ superlattices composed of Mott Insulator-Metal Interfaces (MIMI), realized by alternating layers of $La_2CuO_4$ (LCO) Mott insulator and $La_{1.55}Sr_{0.45}CuO_4$ (LSCO) metallic interfaces [1-7]. The geometric ratio L/d, defined as the superconducting layer thickness over the superlattice period, was finely tuned for each sample, extending both above and below the "magic value" (L/d~2/3) associated with optimal doping, to give the *rising edge* (L/d>2/3) in the normal metal to superconductor transition and drop edge (L/d<2/3) in the superconductor to insulator transition regimes, respectively.

Figure 1 presents the measured superconducting transition temperatures, $T_C$, across all four samples together with theoretical predictions from the BPV theory [1,3], which incorporates the key role of Fano-Feshbach quantum shape resonance between condensates with distinct energy gaps. As discussed in our previous work [5], an enhancement of $T_C$ is observed at optimum lattice geometry L/d~2/3 for the quantum shape resonance, while deviations from this quantum resonance geometry result in reduced $T_C$, at the rising edge and the drop edge of the dome providing strong evidence consistent with two-band superconductivity for the direct link between nanoscale structures and the critical temperature.

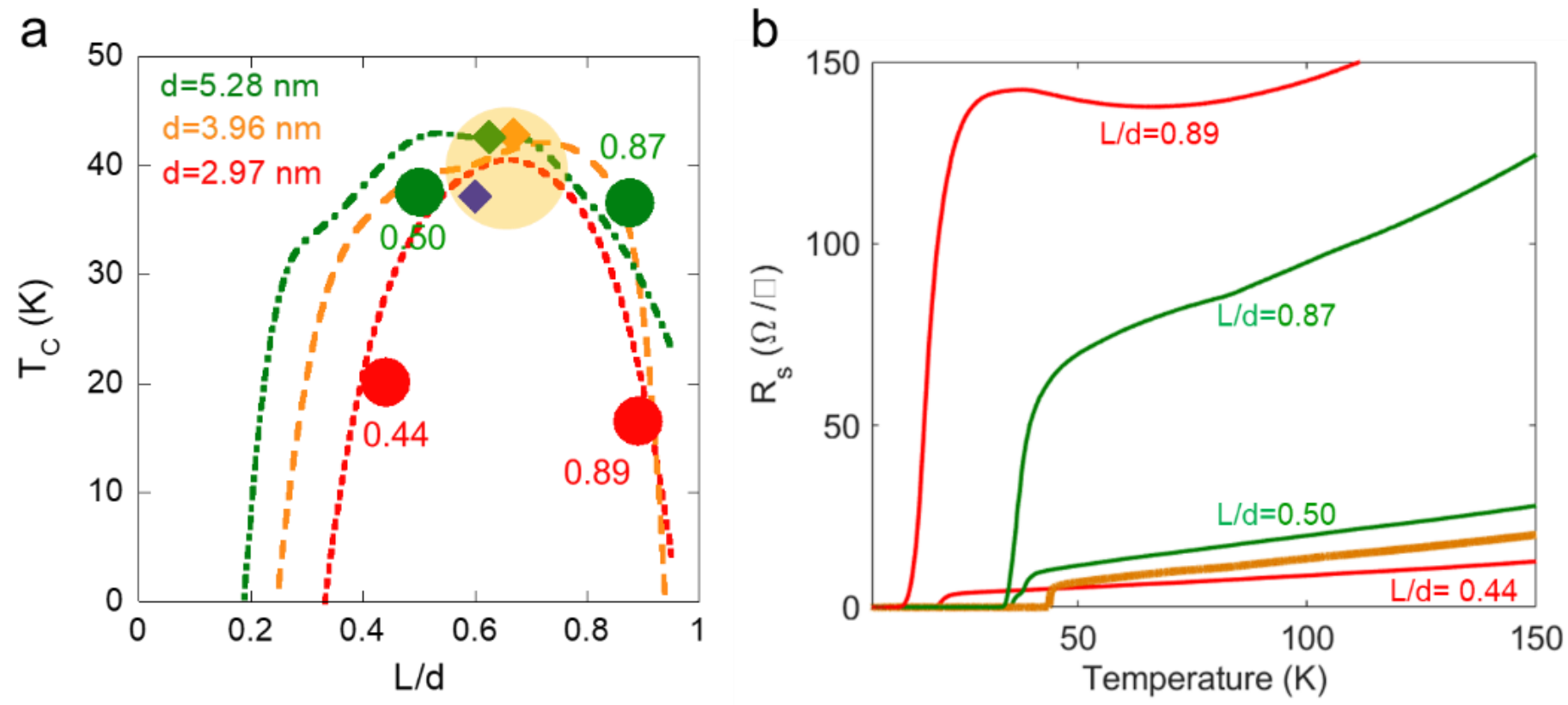

**Figure 1.** (a) Superconducting critical temperature of AHTS superconductors with different L/d values alongside the theoretical curves (dashed lines) calculated by the BPV theory [1,3]. The large full circles represent the four samples studied in this work, alongside to each corresponding L/d value indicated; the diamond symbols represent samples near to the optimum L/d value, studied in [5]. The different colors represent the different unit period, d, as indicated. (b) Sheet resistance as a function of temperature of the AHTS superconductors with different L/d values studied in this work. The sample with $T_C \approx 43$ K around L/d = 2/3, is reported by the thick orange line.

High-field magneto transport measurements (up to 41 T) were conducted at the MagLab DC Field Facility (Tallahassee, FL). Figure 2 reports the resistance $R(\mu_0 H)$ data for samples on the rising edge and on the drop edge a function of temperature and applied magnetic field along the c-axis. The broad field sweep enables us to fully capture the superconducting-to-normal phase transition and to accurately extract critical parameters.

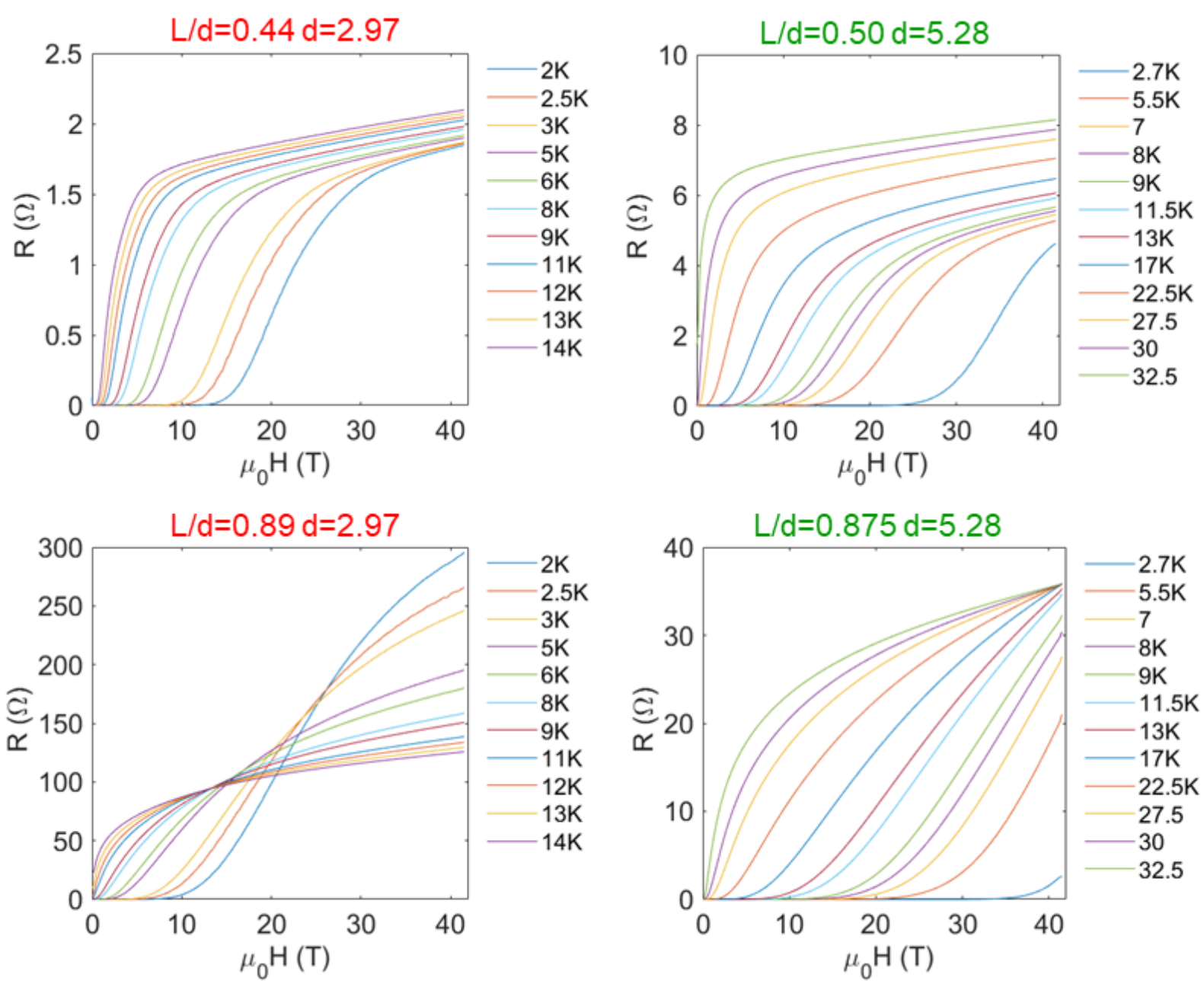

**Figure 2.** Resistance R as a function of the temperature in the four samples with different *L/d* and *d* values indicated, under magnetic field, $\mu_0 H$, measured at MagLab at Florida State University (FSU).

Extraction of the upper critical field ($\mu_0 H_{c2}$) and the irreversibility field ($\mu_0 H_{Cirr}$), shown in Figure 3, and Figure 4 was achieved via both a geometric protocol, and the asymmetric Fano line shape fitting of the derivative $dR/d(\mu_0 H)$, as previously validated in [5]. This robust procedure, applied systematically to all doping regimes, ensures consistent critical field estimates. In Fig. 3a and Fig. 3b we show the derivatives $dR/d(\mu_0 H)$ as a function of $\mu_0 H$ and T for the samples with period d=5.28 and L/d values of 0.50 (low $T_C$) and 0.875 (high $T_C$). We use both linear and logarithmic scales for the temperature, for the sake of clarity. In Figure 3c a further 3D perspective is given by using $dR/d(\mu_0 H)$ as the z-axis for the surface plots.

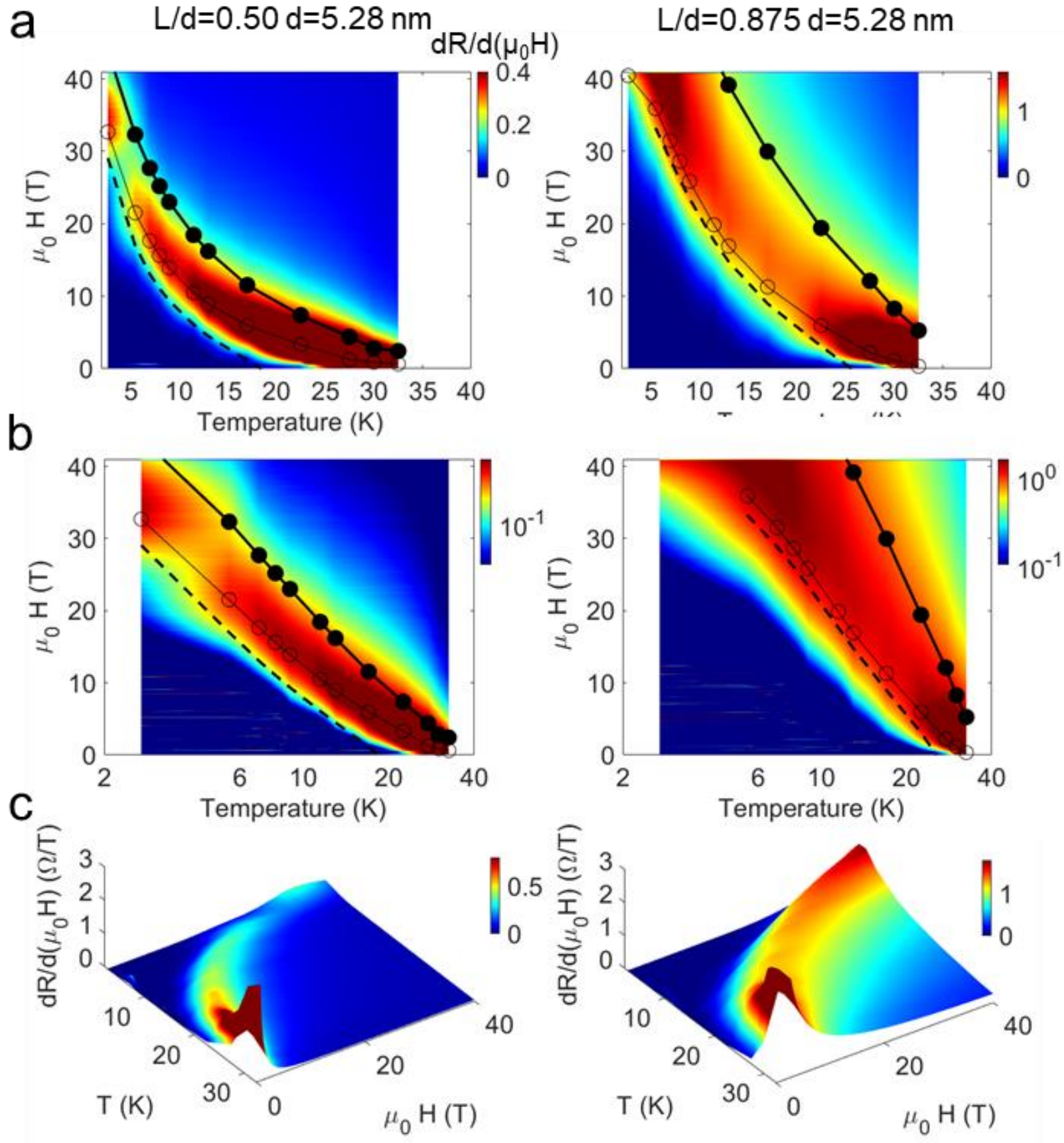

**Figure 3.** (a) Color plots of $dR/d(\mu_0H)$ as a function of $\mu_0H$ and T, for the samples with period d = 5.28 nm with different L/d ratios (0.50 and 0.875). Black dots alongside the thick black line represent the upper critical magnetic fields, $\mu_0H_{c2}(T)$. The dashed black lines are the irreversible magnetic fields $\mu_0H_{Cirr}$. Finally, the empty circles represent the maximum of the derivative, as modelled by the Fano line shape. **(b)** Here, the color plots shown in the (a) panels are plotted in logarithmic scales for both $dR/d(\mu_0H)$ and T. **(c)** 3D perspective of $dR/d(\mu_0H)$ as a function of $\mu_0H$ and T, for both samples.

Similarly, in Fig. 4a and Fig. 4b the derivatives $dR/d(\mu_0H)$ as a function of $\mu_0H$ and T in linear and logarithmic scales as function of the temperature, is shown for the samples with reduced period d=2.97 nm and L/d values of 0.44 (low $T_C$) and 0.89 (high $T_C$). Also, in this case we show a further 3D perspective by setting $dR/d(\mu_0H)$ as z-axis of surface plots in Figure 4c.

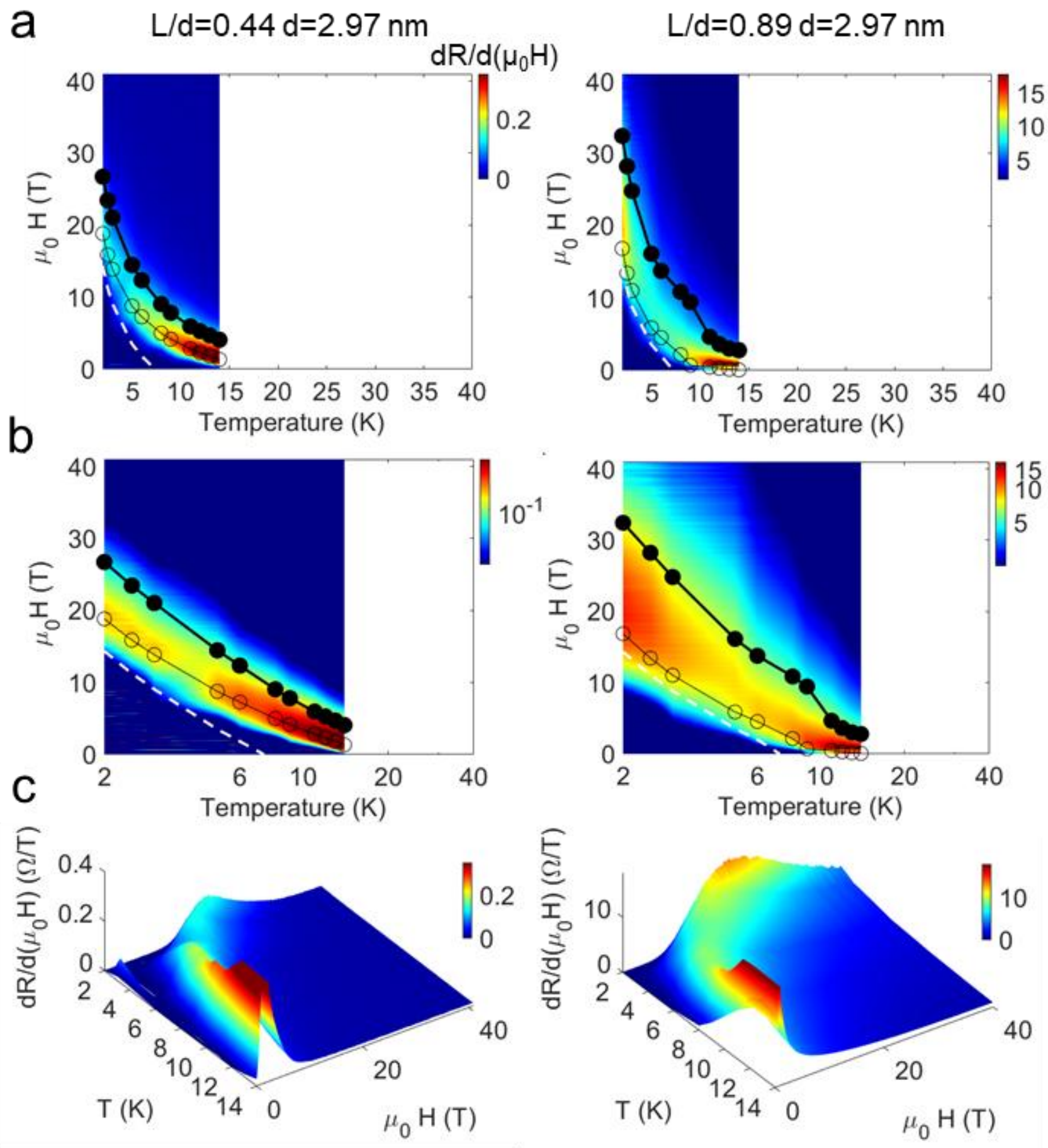

**Figure 4.** Color plots of $dR/d(\mu_0H)$ as a function of $\mu_0H$ and T for the samples with period d = 2.97 nm and L/d ratios of 0.44 and 0.89. Panels show (a) linear and (b) logarithmic temperature scales, and (c) a 3D surface representation. Symbols and lines follow the same conventions as in Fig. 3, which presents the analogous data for samples with d = 2.97 nm and L/d = 0.44 and 0.89.

Upon inspection of $dR/d(\mu_0H)$ we can observe that the zero-temperature upper critical magnetic fields in the AHTS with larger period, d = 5.28 nm, are significantly larger than in AHTS with d = 2.97 nm, in agreement with calculation by BPV theory shown in Figure 1 predicting a quite larger superconducting dome for d=5.28 nm. We observe a quasi-linear dependence of $H_{c2}(T)$ (filled black dots along the solid line) on T in the semi-log plots indicating quasi exponential decrease of $H_{c2}(T)$ as function of T and that the transition becomes broader upon cooling, leading to a faint dR/dH.

This is well depicted in Figure 5a where the upper critical fields, $H_{c2}$, have been plotted as a function of the temperature in the four AHTS. The $\mu_0H_{c2}(T)$ curves of the two samples with lower $T_C$ show a strong upward-concavity providing compelling experimental evidence for robust multigap superconductivity across the entire dome and not just at optimal doping [5]. The sample with L/d=0.5 shows the upward-concave similar to the curve observed for samples near the magic ratio L/d=2/3 [5]. This anomalous upward concavity, contrary to the canonical downward curvature predicted for single-band BCS superconductors [9], demonstrates the key role of two-band physics, in full quantitative agreement with both Ginzburg-Landau two-band theory and Usadel-type diffusive models [5, 13]. A key unexpected result of this study is the particular $\mu_0H_{c2}(T)$ curve observed in the sample with L/d=0.875 in the range 0.66<L/d<0.9 with a very high $H_{c2}(T=0)$ and lower upward-concave curvature indicating a particular pairing regime on the high L/d side of the superconducting dome which is now under further experimental investigation. The systematic evolution of the upward curvature and $\xi_0$ with L/d across the dome need further work on AHTS samples with many L/d ratio across the superconducting dome. The quantitative Usadel-type fits have been reported for AHTS samples near the magic ratio L/d=2/3 [5].

Figure 5b provides a detailed view of the coherence length (ξ) as calculated from the Ginsburg-Landau theory. Indeed, the upper critical field as a function of temperature, $H_{c2}(T)$, can be expressed in terms of the coherence length, ξ(T), using the following relationship derived from Ginsburg-Landau theory:

$$H_{c2}(T) = \frac{\Phi_0}{2\pi\xi(T)^2}$$

where $\Phi_0$ = 2067.8 nm$^2$ T is the magnetic flux quantum. Near $T_C$, the critical divergence is evident, while extrapolations toward zero temperature yield the intrinsic pair size, $\xi_0$, as shown in Figure 5c.

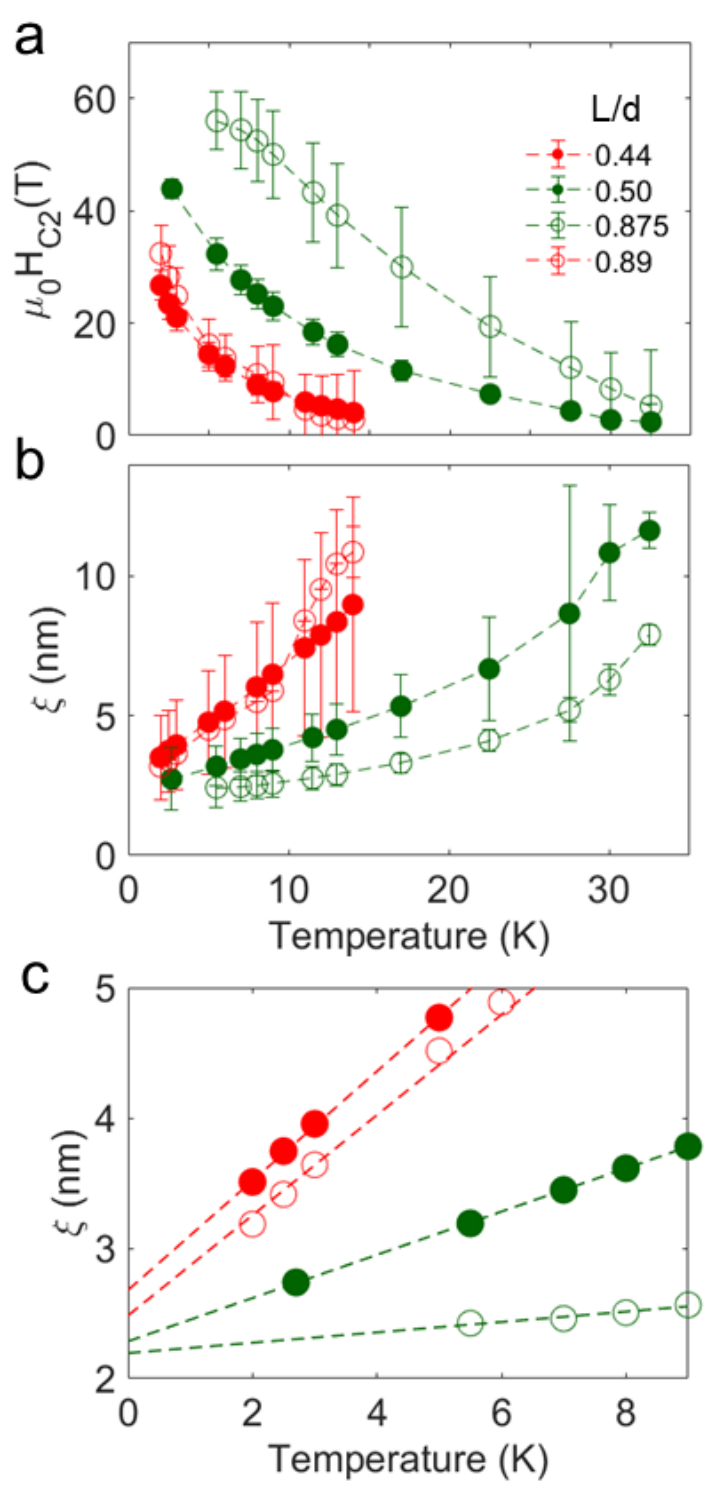

**Figure 5.** (**a**) $\mu_0H_{c2}$(T) as a function of temperature for the four samples studied here (empty circles). For completeness we report also the $\mu_0H_{c2}$(T) values (full circles) obtained for the three samples with L/d= 0.60, 0.625, 0.67, in the optimum doping regime and shown in Campi et al. [5]. Data points at low temperature with $\mu_0H_{c2}$ > 41 T, where the transition becomes broader upon cooling, are extrapolated by the fits of the derivative dR/d($\mu_0$H). (**b**) Temperature dependence of the coherence length. (c) Zoom in the ξ-T plane used to extrapolate the trend toward zero temperature and obtain the intrinsic pair size ($\xi_0$).

A very low coherence length for sample L/d=0.87 in fig. 5c confirms the particularly strong coupling regime for superconductivity on the high L/d border of the superconducting dome. Further higher magnetic field investigations are undergoing at the Los Alamos pulsed facility on samples within the range 0.7<L/d<0.8 where Josephson-junction behavior has been observed in the c-axis transport [8].

Finally, Figure 6 connects the intrinsic pair size, $\xi_0$, across all samples to the geometric parameter L/d and the average doping ⟨δ⟩. The possibility of $T_C$ tunability is revealed, where quantum design of the superlattice architecture can precisely control the fundamental length scale of the Cooper pairs. This correlation provides a predictive roadmap for quantum engineering the next generation of superconductors with tailored condensate sizes.

Further investigations are ongoing to explore new regions of the superconducting dome, aiming to clarify the differences between the dome of the critical temperature and that of the upper critical magnetic field. Notably, the $T_C$ dome as a function of doping in AHTS occurs within the same average hole doping range as natural cuprate oxide perovskites [36]. However, in the drop edge regime of the dome which corresponds to the low-doping regime in natural chemically doped cuprates near the antiferromagnetic to superconducting transition ($\delta \approx 0.05$) AHTS display a sharp increase in $T_C$ which corresponds to the minimum coherence length: this results in a double peak in the $\mu_0 H_{c2}$ versus doping plot, mirroring observations in natural cuprate superconductors such as $YBa_2Cu_3O_y$ [37].

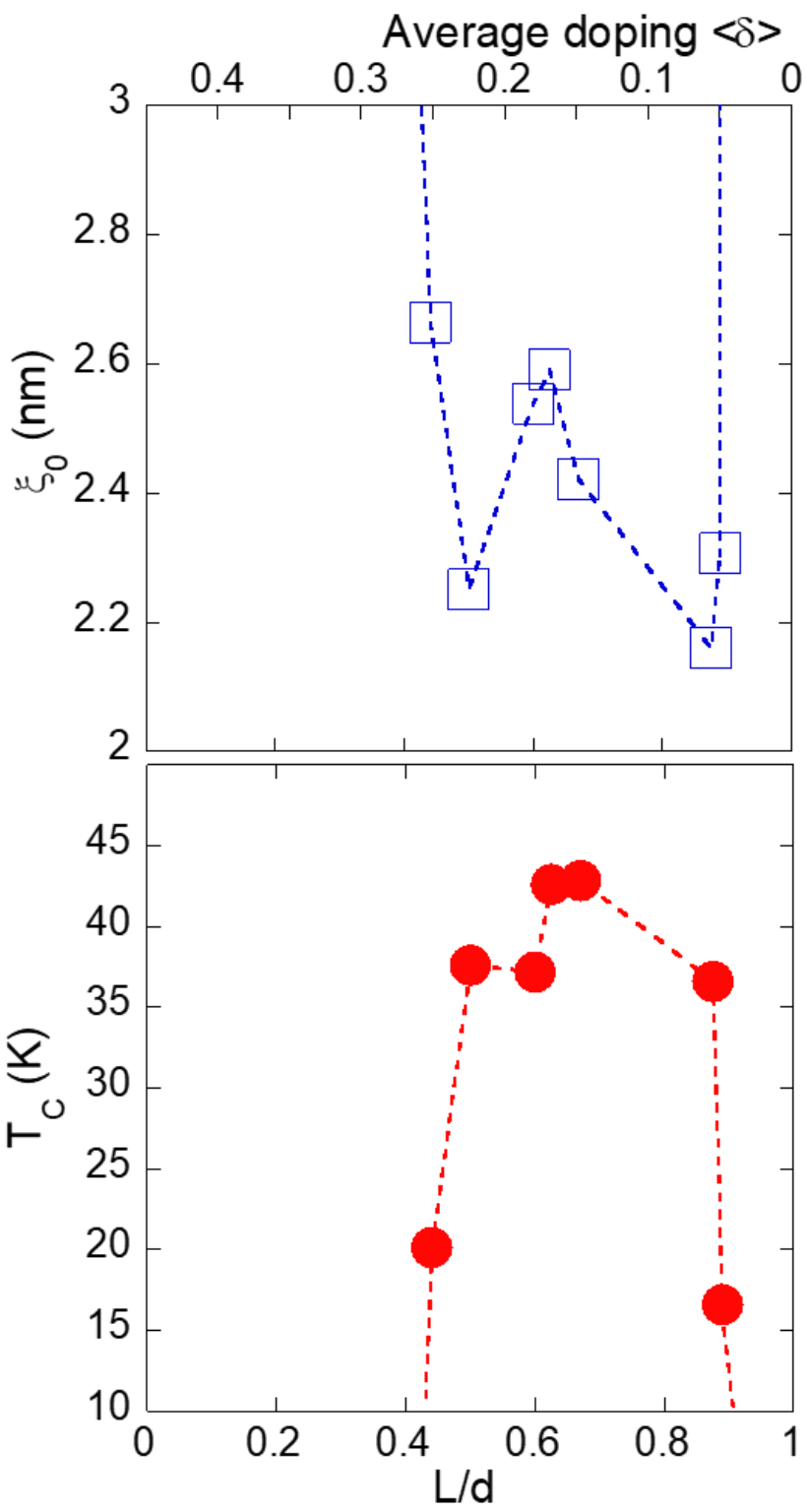

**Figure 6** Cooper pair coherence length, $\xi_0$, and critical temperature, $T_C$, as functions of both the L/d parameter and the average doping $\langle\delta\rangle=0.45(1-L/d)$. The $\xi_0$ values have been extrapolated from the $\xi(T \rightarrow 0K)$ curves highlighted by the black rectangle in panel (c) of Fig. 5. The dashed lines are just guides for the eyes.

The central result of this work, universal upward-concave $H_{c2}(T)$ across the engineered dome, provides direct experimental confirmation that multigap superconductivity persists from dome edges to optimum, as predicted by BPV design. The tunability of intrinsic pair size $\xi_0 \approx 2.2$-$2.6$ nm (Fig. 6) via atomic-scale geometry demonstrates unprecedented control over fundamental superconducting length scales, opening pathways for materials design of next-generation quantum superconducting superlattice.

Unlike Ref. 5 (optimal L/d=2/3 ratio), this comprehensive phase-diagram study establishes: (i) multigap superconductivity also at L/d = 0.44 and 0.89; (ii) a novel high-L/d regime with record-high $H_{c2}(0)$ about 65 T and minimal pair size $\xi_0 \approx 2.2$ nm, suggesting proximity to quantum criticality [Fig. 6]. These findings demonstrate atomic-scale control of both $T_C$ and intrinsic pair size across the full superconducting phase diagram, a capability unique to quantum-designed AHTS.

### III. Conclusions

Our high-field measurements demonstrate that multigap superconductivity is a robust and pervasive feature of Artificial High- $T_C$ Superlattices (AHTS) across the entire doping range investigated. The pronounced upward curvature of the normalized upper critical field $h_{c2}(t)$, observed for all L/d ratios, confirms the coexistence of condensates with distinct Fermi velocities and supports theoretical scenarios based on Fano-Feshbach resonances and BCS-BEC-crossover [6,41] physics governed by quantum size effects tuned by nanoscale superlattice geometry. A key outcome of this study is the emerging difference between the superconducting $T_C$ dome and the $\mu_0 H_{c2}$ dome. While the $T_C$ dome of AHTS spans the similar doping interval as that of natural cuprate perovskites, the low-doping region ($\delta \approx 0.05$) exhibits a sharp enhancement of $T_C$ that coincides with a minimum in the coherence length $\xi_0$. This interplay produces a double-peak structure in the $H_{c2}$ versus doping dependence, as reported in $YBa_2Cu_3O_y$. The divergence between the evolution of $T_C$ and that of $H_{c2}$ highlights the greater sensitivity of the magnetic critical field to multiband pairing and to the tuning of the intrinsic pair size.

These findings suggest that quantum design of superlattice architectures allows not only the modulation of the critical temperature, but also predictive control over the Cooper pair size and upper critical field behavior, opening new perspectives for the next generation of superconducting materials optimized for high magnetic field applications. Ongoing investigations targeting additional regions of the dome will further clarify how quantum design—through structural engineering of L/d, band alignment, and coherence-length tuning—shapes the topology of the $T_C$ and $H_{c2}$ domes, with direct implications for the targeted development of the next-generation of superconducting devices and wires [8,38-40].

## IV Materials and Methods

Artificial high- $T_C$ superlattices (AHTS) were fabricated by molecular beam epitaxy (MBE) employing an ozone-assisted layer-by-layer growth technique. The heterostructures consist of alternating layers of metallic $La_{1.55}Sr_{0.45}CuO_4$ (LSCO) and $La_2CuO_4$ (LCO), with the latter hosting the superconducting space-charge regions. Growth was carried out on $LaSrAlO_4$ (001) substrates, resulting in a compressive lattice mismatch for LCO of approximately +1.4%. The deposition process was performed with an ultra-low kinetic-energy flux (~0.1 eV), and the assembly of each monolayer was monitored in real time using reflection high-energy electron diffraction (RHEED). The substrate temperature, measured by a calibrated pyrometer, was held at 650 °C. The chamber pressure during deposition was maintained at approximately $1.5\times10^{-5}$ Torr in an environment containing ozone, atomic, and molecular oxygen. Upon completion of the growth sequence, samples were cooled to 200 °C prior to switching off the ozone input and subsequently cooled to room temperature under high vacuum conditions to eliminate interstitial oxygen from the LCO layers and restore stoichiometric composition. Structural integrity and superlattice periodicity were verified by synchrotron x-ray diffraction at the Elettra facility, Trieste [42].

Resistance measurements were performed by using a four-point van der Pauw configuration with alternating DC currents of ±10 μA, over a temperature range from room temperature down to 4.2 K (liquid helium). We used a motorized, custom-made dipstick inserted in a transport helium dewar, ensuring a temperature variation rate below 0.1 K/s.

Magneto transport measurements were conducted at the National High Magnetic Field Laboratory (NHMFL), Florida State University. The upper critical field was determined using a resistive Bitter magnet capable of generating static magnetic fields up to 41 T, combined with a variable-temperature insert enabling continuous temperature control from 1.30 K to 35 K via calibrated Cernox sensors. Sample orientation with respect to the magnetic field was adjusted using a single-axis rotator with a Hall probe for precise angular alignment. Longitudinal resistance was measured in the van der Pauw geometry using gold pads thermally evaporated onto the substrate corners, with four-probe detection provided by a Lakeshore 370 resistance bridge. Data analysis was performed by using MATLAB routines enclosed in homemade package [43].

## Acknowledgements

We thank the Superstripes-onlus and the CNR project DCM.AD006.562 “Functional Disorder in Materials and Complex Systems” for supporting this work. L.B. acknowledges support from the US DoE, BES program, through award DE-SC0002613. A portion of this work was performed at the National High Magnetic Field Laboratory, which is supported by National Science Foundation Cooperative Agreement No. DMR-2128556 and the State of Florida.